\documentclass[journal]{IEEEtran}
\usepackage[figuresright]{rotating}
\usepackage{graphicx}
\usepackage{subfigure}
\usepackage{cite}
\usepackage{multirow}
\usepackage{tabularx}
\usepackage{caption}
\usepackage{longtable}
\usepackage{makecell}
\usepackage{lscape}
\usepackage{float}
\usepackage{threeparttable}
\usepackage{booktabs}

\begin{document}
%
\title{A Systematic Survey of Blockchained Federated Learning}



%
%
%

\author{Zhilin Wang, Qin Hu, Minghui Xu, Yan Zhuang, Yawei Wang, Xiuzhen Cheng
\thanks{Zhilin Wang is with the Department of Computer Science, Purdue University, West Lafayette, Indiana, USA. E-mail: wang5327@purdue.edu}

\thanks{Qin Hu is with the Department of Computer Science, Indiana University Indianapolis, Indiana, USA. E-mail: qinhu@iu.edu}

\thanks{Yan Zhuang is with the Department of BioHealth Informatics, Indiana University Indianapolis, Indiana, USA, Email: ynzhang@iu.edu}

\thanks{Yawei Wang is with the Department of Computer Science, The George Washington University, Washington, DC, USA. Email: yawei@gwu.edu}

\thanks{Minghui Xu and Xiuzhen Cheng are with the School of Computer Science and Technology, Shandong University, Qingdao, Shangdong, China. Email:\{mhxu,xzcheng\}@sdu.edu.cn }
}

\maketitle

\begin{abstract}
With the technological advances in machine learning, effective ways are available to process the huge amount of data generated in real life. While the issues of privacy and scalability will constrain the development of machine learning. Fortunately, federated learning (FL) can prevent privacy leakage by assigning training tasks to multiple clients, thus separating the central server from local devices. However, FL still suffers from shortcomings such as single-point-failure and malicious data. The emergence of blockchain provides a secure and efficient solution for the deployment of FL. In this paper, we conduct a comprehensive survey of the literature on blockchained FL (BCFL). First, we investigate how blockchain can be applied to FL from the perspective of system composition. To the best of our knowledge, it's the first paper to classify the combination of FL and blockchain into fully coupled BCFL, flexibly coupled BCFL, and loosely coupled BCFL based on the coupling level.
Then, we analyze the concrete functions of BCFL from the perspective of mechanism design and illustrate the problems blockchain addresses specifically for FL. Next, we survey the applications of BCFL in reality. Finally, we discuss the deficiencies of BCFL, i.e., the new challenges that the inclusion of blockchain brings to FL, and some future research directions are also discussed. We hope this paper can bring some insightful ideas to future research on BCFL.
\end{abstract}


%
\IEEEpeerreviewmaketitle

\maketitle

\section{Introduction} \label{sec:intro}


Nowadays, the applications of machine learning (ML) in different fields have profoundly changed human life. Daily generated data can be gathered from massive end users to train ML models which can bring benefits in terms of providing better services to improve our quality of life. However, 
the current ML framework usually requires end devices to transfer the collected data to the central server for model training, thus leading to two challenges: 1) data transferring can consume a large amount of communication resources, and 2) the submission of raw data increases the risk of privacy leakage, making data owners reluctant to upload data to the central server for the security concern.


To address above challenges, Google proposed a novel ML framework, named federated learning (FL), to effectively protect the privacy of data owners, i.e., clients, while collaboratively training an ML model in a distributed manner \cite{mcmahan2017communication}. Different from the conventional ML framework, FL does not require data owners to transfer the raw data to the central server for model training; instead, only the parameters of local models trained using the local data need to be submitted for global model aggregation. By this means, both the massive data transmission cost and potential privacy leakage caused by transferring the original data can be largely alleviated. In the past few years, FL has been studied extensively and developed rapidly \cite{Bonawitz2019, Hard2018}. However, the traditional FL framework still faces some problems which undermine the performance and reliability of the whole system \cite{Sattler2020, mcmahan2017communication, Bagdasaryan2018}, which can be summarized as follows.

\begin{itemize}
    \item \textbf{Single point of failure.} In FL, a central parameter server, which is also named as \textit{aggregator}, is employed to perform the integration of local training results so as to update the global model. However, the aggregator is not always reliable. Once the centralized aggregator is compromised, the whole FL system can easily come into a failure. Some potential problems of the aggregator include intentionally dishonest aggregation, accidental network connection failure,  unexpected external attacks, etc.
    
    \item \textbf{Malicious clients and invalid submissions.} Given the large number of participants in FL, it might be impractical to assume that all clients are honest and participate in FL according to the predefined protocols. There may exist malicious clients submitting invalid data about their local training results, heavily impacting the performance of the global model. 
    And the whole FL system might also be undermined by selfish or lazy clients via other means, such as training the local models using partial datasets.
    
    \item \textbf{The lack of incentives.} In the traditional FL, it is usually considered that all clients willingly contribute their computing power and local data. Nevertheless, this ideal scenario without any payments to clients can make it difficult to motivate them to honestly follow the protocol regarding complete calculation and reliable data provision. More importantly, since FL usually requires multiple devices to work collaboratively, especially for large-scale data-intensive training tasks in need of wide participation, the traditional FL framework may fail to attract enough number of clients engaging in the FL training due to the lack of incentives.
    
\end{itemize}

To address the above challenges, scholars have tried a variety of methods for achieving higher efficiency and reliability of the conventional FL. 
Among them, blockchain, as an emerging technology with several attractive properties, such as decentralization, anonymity, and traceability, has been widely employed in this field to enable the full distribution property of FL, exerting exemplary performance, which can be termed as blockchained FL (BCFL) to differentiate from the traditional FL.
Technically, blockchain can provide the following representative functions to enhance FL. 
First, full decentralization can be realized by deploying blockchain in FL, where the central aggregator can be replaced by the peer-to-peer blockchain system to handle the job of updating the global model, thus alleviating the vulnerability of the whole FL system caused by the failure of the centralized server \cite{Ramanan2019}. 
Second, blockchain can facilitate the verification of model updates for FL in the process of transaction verification, by which the low-quality or even malicious local submissions can be removed before the global model is aggregated \cite{kim2019blockchain}.
In addition, blockchain can enable efficient reward distribution to FL clients for encouraging their participation and honest behaviors \cite{Liu2020}. 

Based on our investigation, we argue that blockchained FL framework has at least the following merits:

\begin{itemize}
    \item Single point of failure can be avoided when the central aggregator is taken place by the blockchain system, where the model aggregation will be executed on the chain.

    \item Unreliable submissions from clients can be filtered out with the help of verification mechanisms in blockchain. Before local model updates are aggregated to update the global model, abnormal data can be detected, and only valid data will be employed to calculate the global model.
    
    \item More participants and computational resources can be attracted through the incentive mechanisms implemented on the blockchain. Several types of economical incentives, such as cryptocurrency, can be utilized to encourage more clients to participate in the model training for data-intensive FL tasks and drive them to behave legitimately regarding following the predefined FL rules.
    
    \item Learning data can be permanently stored and effectively shared on a distributed ledger. Once the data are recorded on the main chain, it becomes hard for them being tampered. Meanwhile, authorized clients can get access to the blockchain to retrieve the historical data regarding stored on blockchain, facilitating data queries.
\end{itemize}

Despite the excellent performance of BCFL in terms of achieving decentralization and providing incentives, several new problems, such as resource allocation, communication delays, and external attacks, arising from the combination of the two techniques still need to be well addressed in future research. Besides, the major problem with current research on BCFL is lacking adequate understanding of how blockchain and FL can be consolidated to achieve better efficiency and security performance.

There are few surveys exploring the topic of BCFL. In \cite{li2021blockchain}, the authors briefly summarize the latest progress in BCFL and discuss how to improve the performance of BCFL. The work in \cite{li2021blockchain2} analyse BCFL from the perspective of its layers composition. However, all of them are mainly a simple combination of blockchain and
FL, lacking systematic analysis and in-depth discussion of the new challenges faced by BCFL.

To fill the gap, we carry out this survey in a comprehensive way.
We first introduce the background of FL and blockchain, and then we survey the architectures, blockchain types, and learning devices of BCFL. Next, we illustrate the four main functions of BCFL, i.e., the verification of model updates, the aggregation of global model, the utilization of the distributed ledger, and incentive mechanisms. We also analyze the applications of BCFL in the Internet of Things (IoT), healthcare, and business. In addition, we discuss the new challenges of BCFL and explore some promising future research directions.

To the best of our knowledge, this paper is the first comprehensive investigation of BCFL. Followings are the main contributions of our work:

\begin{itemize}

    \item We investigate the current research of blockchained FL, and analyze the motivations of applying blockchain to FL.
    
    \item We detail the foundations of BCFL, including the BCFL architecture, blockchain types, and training devices. We are the first to propose that BCFL architectures can be classified into three types according to the relationship between FL clients and blockchain nodes: fully coupled BCFL, flexibly coupled BCFL, and loosely coupled BCFL.
    
    \item We present the functions of BCFL from the perspectives of verification mechanism, global model aggregation, distributed ledger and incentive mechanism. The analysis of these functions explains the changes that blockchain can bring to FL.
    
    \item We analyze the current challenges of BCFL and discuss the future research directions.
    
\end{itemize}

The rest of this article is organized as follows. In Section \ref{sec:basics}, we introduce the basics of FL and blockchain. We present the foundations of BCFL in Section \ref{sec:founda}. In Section \ref{sec:functions}, we detail the four functions of BCFL. And in Section \ref{sec:applications}, we investigate the applications of BCFL in different domains. Discussions of the current challenges and future research directions of BCFL are presented in Section \ref{sec:challenges}. Finally, we conclude the paper in Section \ref{sec:conclusion}. For convenience, major abbreviations used in this paper are summarized in Table \ref{table:abbre}.

\begin{table}
\centering
\caption{Terminologies.}
\label{table:abbre}
\begin{tabular}{|c|c|c|}
\hline
  Abbreviation   & Meaning                                          \\ \hline \hline
 BC      & blockchain                                       \\ \hline
 FL      & federated learning                               \\ \hline
 BCFL    & blockchained FL              \\ \hline
 PoW     & proof of work                                    \\ \hline
PoS     & proof of stake                                   \\ \hline
 SC      & smart contract                                   \\ \hline
BFT     & Byzatine fault tolerant                          \\ \hline
FuC-BCFL & fully coupled blockchained FL    \\ \hline
 FlC-BCFL & flexibly coupled blockchained FL \\ \hline

 LoC-BCFL & loosely coupled blockchained FL \\ \hline

\end{tabular}
\end{table}


\section{Background Knowledge}\label{sec:basics}
In this section, we will go through the basic principles of blockchain and FL to prepare readers for understanding BCFL. 


\subsection{Brief Introduction to FL}

In real life, mobile devices with various embedded sensors are used extensively, generating a massive amount of data. However, traditional ML frameworks are unable to effectively process such large amounts of distributed data due to the following three challenges.  
First, the data generated by multiple devices are usually unbalanced and non-independent and identically distributed (non-IID). Moreover, communication cost among devices is expensive since they are massively distributed \cite{Zhao2018, Yang2020, Kairouz2019}. In addition, storing all the data in a centralized manner is not a secure choice. Therefore, Google introduced a novel distributed machine learning framework termed FL to address the above issues of applying traditional machine learning on mobile devices \cite{konevcny2016federated, mcmahan2017communication, Konecny2016}. 

\begin{figure}[h]
\centering
\includegraphics[width=8cm]{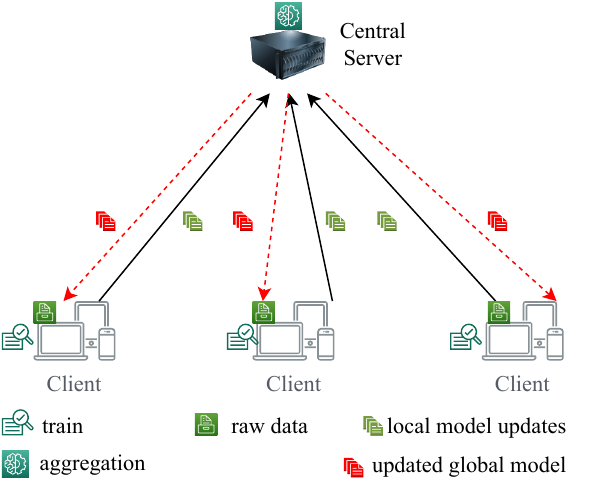}
\caption{The topology of FL.}
\label{fig:tra_fl}
\end{figure}

FL trains data on local devices in a decentralized manner, local devices then upload local model updates, i. e., weights or weight differences of the local models, to a central server, which will run a predefined aggregation algorithm to update the global model. The topology of FL is shown in Fig. \ref{fig:tra_fl}. Usually, local devices are named as \textit{clients}, and the central server is termed as \textit{aggregator}. The basic merit of FL is that it requires no access to the raw data on local devices directly \cite{mcmahan2017communication}. In this way, FL allows the privacy of the raw data to be preserved effectively and also reduces the cost of data transmission, which makes it more applicable to mobile devices \cite{Li2019}. The workflow of typical FL is described as follows\cite{Konecny2016, Hard2018, Nishio2019}: 

\begin{itemize}
    \item \textbf{Client Selection:} Clients are selected based on definite protocols, and then they download the latest global model before the training task starts.
    
    \item \textbf{Local Model Training:} Clients train local models based on predefined algorithms (e.g., Stochastic Gradient Decent (SGD)) independently using their local data.
    
    \item \textbf{Upload Local Model Updates:} Clients transfer local model updates to the aggregator.
    
    \item \textbf{Global Model Aggregation:} The global model is calculated by the aggregator through executing the aggregation algorithm, such as FederatedAveraging (FedAvg).
    
\end{itemize}

FL can be classified into three categories, i.e., Horizontal FL, Vertical FL and Federated Transfer Learning, based on the distribution characteristics of raw data, which can be detailed as below \cite{Yang2019}.
\begin{itemize}
    \item \textbf{Horizontal FL:} The datasets owned by FL clients have the same feature space with varying samples. 
    
    \item \textbf{Vertical FL:} The datasets of clients  have the same sample space while the feature space is various.
    
    \item \textbf{Federated Transfer Learning:} In the case where there are two datasets have less overlapping samples and features, transfer learning is used to overcome the lack of data or labels without slicing the data.
\end{itemize}





Though FL has the advantages, such as protecting the privacy of raw data and reducing the communication cost to some extent, it still faces some challenges, i.e., security issues, privacy leakage of local model updates, and the cost of multi-rounds communication between clients and central server, etc. In order to improve the stability and sustainability of FL, many studies have been conducted. 


As for the security and privacy protection of local models, the work in \cite{Mothukuri2021} surveys the research related to the privacy issues of FL, illustrating several attacks that can lead to the leakage of data privacy, e.g., membership inference attacks and Generative Adversarial Nets (GAN) based inference attacks. Meanwhile, the authors introduce several countermeasures based on Differential Privacy (DP) and Secure Multi-party Computation (SMC). DP is a commonly applied technology which preserves data privacy by adding noises to private information, i.e., local model updates in FL that are required to be uploaded to the central server. DP reduces the possibility of the raw data being inferred reversely without significant loss of data quality. Nowadays, DP is widely used in FL to fully protect the privacy of clients \cite{Geyer2017, Choudhury2019, Truex2019}.

To address the concern of multi-round communication cost between clients and the server, Sattler et al \cite{Sattler2020} argue that the methodology of collaborative training protects data privacy, but it causes communication challenges, e.g., the increase in communication cost. They propose a novel compression algorithm termed Sparse Ternary Compression (STC) to address the above issue. The experimental results indicate that STC is effective in general situations. Besides, a communication-efficient Federated Learning Method with Periodic Averaging and
Quantization (FedPAQ) is another methodology designed to overcome the challenges of communication bottleneck and scalability while guaranteeing the accuracy of FL in the meantime~\cite{Reisizadeh2019}.


FedPAQ allows partial nodes to participate in the local training, and then nodes transfer the qualified updates to the parameter server which averages the global model periodically.

Since FL framework was proposed by Google in 2016, it has been used has  been  used  practically in  many  areas, including wireless computing\cite{Chen2020a, Niknam2020, li2020review}, healthcare\cite{xu2019federated, brisimi2018federated, chen2020fedhealth}, Internet of Things\cite{Du2020, khan2020federated}, smart city\cite{qolomany2020particle, jiang2020federated}, business and finance~\cite{Yang2019}.


    
    



















\subsection{Brief Introduction to Blockchain}
Blockchain technology originated from Bitcoin proposed by Satoshi Nakamoto in 2008~\cite{nakamoto2019bitcoin}, which has attracted tremendous interest from government and academia to industry for its decentralization, transparency, and immutability. 
\begin{itemize}
\item \textbf{Decentralization:} The decentralized nature of the blockchain solves the drawbacks of traditional system centralization and prevents single points of failure, and its decentralization is reflected in the decentralized consensus, incentive mechanism, and data storage. In the blockchain, the attacks on any node can not destroy the operation of the entire system, greatly improving the security and stability of the system.
\item \textbf{Transparency:} The transparency of the blockchain mainly refers to the data on the chain being visible to available partners. Every legal user can check all transactions in any block according to the transaction address or transaction ID. Moreover, due to the anonymity of the blockchain, it does not reveal the user's identity privacy while being transparent.
\item \textbf{Immutability:} The immutability of the blockchain means that once the user completes the transaction, his record will be permanently stored on the blockchain ledger, and no one can change it. The bottom layer uses the Merkle tree structure to verify the immutable data. In addition, the blockchain still uses technologies such as digital signatures, hash calculations, and time stamps to ensure that the data stored in the blockchain cannot be tampered with, increasing the credibility of the system.
\end{itemize}

Due to the above properties, blockchains can solve the traditional single point of failure problem, achieve effective access control, and even provide a zero-trust fault-tolerant environment, enabling mutually distrustful parties to establish a trust relationship in wireless networks.
Hence, the blockchain is quickly introduced to many areas including cryptocurrency~\cite{wood2014ethereum}, healthcare~\cite{agbo2019blockchain, tanwar2020blockchain}, smart city~\cite{Xie2019, aujla2020blocksdn} and Internet of Things (IoT)~\cite{wang2020blockchain}. 



Blockchain is essentially a distributed database that is empowered by miners. Each miner keeps one replica of the entire ledger locally and competes to obtain certain rewards by packaging verified transactions.
Typically, blockchain can be roughly classified into three categories, i.e., public blockchain, private blockchain, and consortium blockchain. These three categories differ in the restrictions on the members who can participate. To be specific, in the public blockchain, everyone can join and leave without permission. The consortium blockchain is partially decentralized and controlled by several predefined or selected users (i.e., authorities who have the right to generate new blocks). 
In the private blockchain, users are under supervision and the entire system is maintained by a person or an organization. The specific differences are shown in Table \ref{table:comparation}.
\begin{table}[htbp!]
		\centering
		\caption{Comparison of different types of blockchains.}
		\label{table:comparation}
		\begin{tabular}{c c c c c }
			\toprule
x			Blockchain   & Participants   & Decentralization &   Access control \\ 
			\specialrule{0em}{1pt}{1pt}
			\midrule
			Public &  All users     & Completely &  No \\
			\specialrule{0em}{1pt}{1pt}
			Consortium  & Partial users   & Partially & Yes \\
			\specialrule{0em}{1pt}{1pt}
			Private  & Partial users   & Partially & Yes \\
			\bottomrule
		\end{tabular}
\end{table}

\section{Foundations of BCFL}\label{sec:founda}

In this paper, we investigate the implementation of several new features in the FL model through blockchain, so as to address some existing problems of FL. In this section, we explore BCFL as a whole system, describing and classifying its architectures. Our work is based on the perspective of the components of the BCFL model. At the beginning, we propose a methodology to classify the architectures of BCFL according to the coupling between blockchain and FL. Next, we analyze blockchain and FL in 
BCFL system respectively. Since the blockchain has different types, various properties that BCFL models have on different types of chains are discussed. We notice that the participants of the model training of BCFL are distinct, which will affect the deployment of BCFL in specific applications. We will also provide lessons learned in each subsection to illustrate more concretely how BCFL model works. Table \ref{table:summery} shows the summery of the relevant literatures.

\begin{sidewaystable*}
\center
\tabcolsep 4pt 
\renewcommand\arraystretch{1.3} 

\caption{Table of the literature related to blockchained FL}
\scriptsize 
\begin{tabular}{ccm{80pt}<{\raggedright}m{40pt}<{\raggedright}m{60pt}<{\raggedright}
m{40pt}<{\raggedright}m{70pt}<{\raggedright}m{280pt}<{\raggedright}}
  \hline\\[-2.6mm]\hline
 &	Ref. & Blockchain & Learning devices & Blockchain types & Consensus protocol & Central aggregator	& Contribution\\
 \hline
\multirow{15}{*}{\makecell{Fully coupled}}
 & \cite{Lu2020c} & / & Device & Permissioned & PoQ & No & Blockchained FL to share and retrieve data for IoT devices\\
  & \cite{cao2021towards} & / & Device & Public & DAG-FL & No &Introduced a direct acyclic graph based FL consensus (DAG-FL) to address the asynchrony of devices and anomaly detection of BCFL\\
 & \cite{Mugunthan2020} & Ethereum & / & Public & / & No & BlockFlow is an accountable FL system that is fully decentraized and privacy-preserving\\
 & \cite{Preuveneers2018} & MultiChain & / & Permissioned & PoC & No &  Proposed a permissioned blockchained FL to do anomaly detection\\
 & \cite{Li2020c} & / & / & Permissioned & CCM & No & Proposed BFLC which defines the model storage pattens, the training process and committee consensus\\
 & \cite{Ramanan2019} & Ethereum & / & Public/ permissioned & PoA & No & Leveraged smart contract to take place of the central aggregator\\
 & \cite{Toyoda2019} & Ethereum & / & Public & / & No & Designed a competitive incentive mechanism\\
 & \cite{kim2019blockchain} & / & Edge & Public & / & No & Proposed node recognition based local learning weighting method and node selection method\\
 & \cite{Hua2020} & / & / & / & / & No & Used blockchain to store, transfer and share machine learning models\\
 & \cite{Chai2020} & / & Device & Public & PoK & No & Hierarchical BCFL to share information in IoV\\
 & \cite{weng2019deepchain} & / & / & public & blockwise-BA & No & DeepChain provides a value-driven incentive mechanism based on BC\\
 & \cite{Bao2019} & / & / & Public & / & No & Proposed FLChain to distribute trust and incentive among trainers\\
 \hline
\multirow{30}{*}{\makecell{Flexibly coupled}}
 & \cite{Qu2020} & / & Device & / & PoW & Yes & Prevented single point failure in fog computing based on FL-Block\\
 & \cite{Majeed2019} & Ethereum & Edge & / & PoW & No & Leveraged channels to train models and global model state of tie to aggregate model\\
 & \cite{Kim2019} & Ethereum & Device & Public & PoW & No & Blockchained FL to exchange and verify model updates and latency analysis\\
 & \cite{passerat2019blockchain}  & Ethereum & / & Permissioned & / & Yes & Privacy-preserved healthcare consortia based on BCFL\\
 & \cite{Hieu2020} & / & Device & / & PoW & 	Yes & Used deep reinforcement learning to derive the optimal decisions for the model owner\\
 & \cite{Lu2020b} & / & Edge & Permissioned & DPoS & Yes & Leveraged blockchain to train local model updates before global aggregation\\
 & \cite{Li2020b} & Ethereum & / & Permissioned & PoW/PoS & Yes & Transplanted the entire crowdsourcing system onto the blockchain\\
 & \cite{Ma2020} & / & Device & Public & PoW & No & The central server was replaced by blockchain\\
 & \cite{Lu2020} & / & Edge & Permissioned & DPoS & No & Developed a lightweight verification scheme for permissioned blockchain based on DPoS\\
 & \cite{Pokhrel2020} & / & Device & / & PoW & No & Autonomous vehicular system based on BCFL\\
& \cite{desai2021blockfla} & Hyperledger fabric/ethereum & Device& Permissioned/ public & / & Yes & Proposed a BCFL to deter adversarial attacks by accounting\\
 & \cite{Zhao2020} & / & / & Permissioned & PoS/PoQ & No & Used the blockchain to replace the central aggregator\\
 & \cite{Martinez2019} & EOS & Device & Public & PoC & Yes & Using EOS BC and IPFS to record uploaded updates in a scalable manner and reward users based on training data cost\\
 & \cite{Lu2020a} & / & Edge & Permissioned & DPoS & Yes & Developed hybrid blockchain and asynchronous FL to share data in IoT\\
 & \cite{Sharma2020} & / & Edge & / & / & Yes & Proposed a multi-layer distributed computing defence framework\\
 & \cite{Shen2020} & / & Edge & Permissioned & / & Yes & analyzed the unintended property leakage in BCFL for intelligent edge computing\\
 & \cite{Zhang2020a} & Ethereum & / & / & PoW & Yes & Designed an anchoring protocol to build a Merkle tree\\
 & \cite{Cui2020} & Ethereum & Edge & Permissioned & PoS & Yes & Blockchain serves as a distributed ledger that records the transactions in terms of models and training parameters\\
 & \cite{Liu2020} & Ethereum & Device & / & / & Yes & Prevented poisoning and membership inference attacks\\
 \hline
\multirow{10}{*}{\makecell{Loosely coupled}}
 & \cite{UrRehman2020} & Ethereum & Edge & Public & PoW & Yes & Designed a blockchain based reputation-aware fine-gained FL\\
 & \cite{kang2019incentive} & / & Edge & Permissioned & / & Yes & Reputation management and incentive mechanism based on blockchain\\
 & \cite{kumar2021blockchain} & / & / & Public & / & Yes & Leveraged blockchain to retrieve data in hospitals\\
 & \cite{Kang2020} & / & Edge & Permissioned & / & Yes & The concept of reputation is introduced as a metric\\
 & \cite{fan2020hybrid} & Ethereum & Edge & Public/ permissioned & PBFT & Yes & Leveraged the hybrid blockchained system for FL in edge computing\\
 & \cite{Doku2020} & / & Edge & / & PoCI & No & Ensured the data used to train models in the network is trustworthy and relevant\\
 & \cite{Nagar2019}& Ethereum & Device & Public & PoW & Yes & The blockchain network enables exchanging device's local model updates while verifying their work\\
  \hline\\[-2.6mm]\hline
  &  &  &  &  &  &  & / means that the relevant information is not clearly mentioned in that literature.\\
	
\label{table:summery}
\end{tabular}
\end{sidewaystable*}

For rigorous expression in this paper, some terminologies of blockchained FL are listed and explained below:
\begin{itemize}
    \item \textbf{Clients:} devices that work in FL system to collect data and train local models.
    \item \textbf{Nodes:} members in blockchain network to provide computing powers and generate new blocks, which can also be called miners.
    \item \textbf{Aggregator:} server or other powerful enough equipments to aggregate the global model.
    \item \textbf{Distributed ledger:}  a traceable and audible database distributed across multiple nodes in blockchain network, storing data for retrieve or audit.
    \item \textbf{Transaction:} data records in each block.
    \item \textbf{Local Model updates:} gradients and weights computed by clients based on local raw data.
    \item \textbf{Rewards:} Rewards can be from blockchain system or FL system based on certain evaluation method. 
\end{itemize}

\subsection{Architectures of BCFL}\label{sec:framwork}

Before we design the BCFL model, a clear understanding of its architecture is necessary; however, no relevant studies have been conducted on the architecture of BCFL. In our paper, we will fill this gap. Inspired by \cite{hu2021blockchain}, we classify the architectures of BCFL into three categories based on different coupling: fully coupled BCFL, flexibly coupled BCFL and loosely coupled BCFL.

\subsubsection{Fully Coupled BCFL}

We can define the the framework as the fully coupled blockchained FL  model (FuC-BCFL) when the clients of FL are the nodes of blockchain, in other words, the clients not only train the local models, but also verify the updates and generate new blocks. The topology of FuC-BCFL is shown in Fig. \ref{fig:fully}. We can derived from the definition of FuC-BCFL that FL model is decentralized 
since every node on blockchain has the chance to participate in the local model training and global model aggregation, thus the role of central aggregator can be taken place by the blockchain. In such a framework, there are two methodologies to average the global model: i) some selected nodes collect the validated local model updates and then conduct the aggregation algorithm; ii) all the nodes can participate in the global model aggregation. The distributed ledger contains the training data, including the verified local model updates, global model updates and other data produced during the learning process.  Typically, the workflow of the FuC-BCFL can be summarized as follows:

\begin{figure}[h]
\centering
\includegraphics[width=8cm]{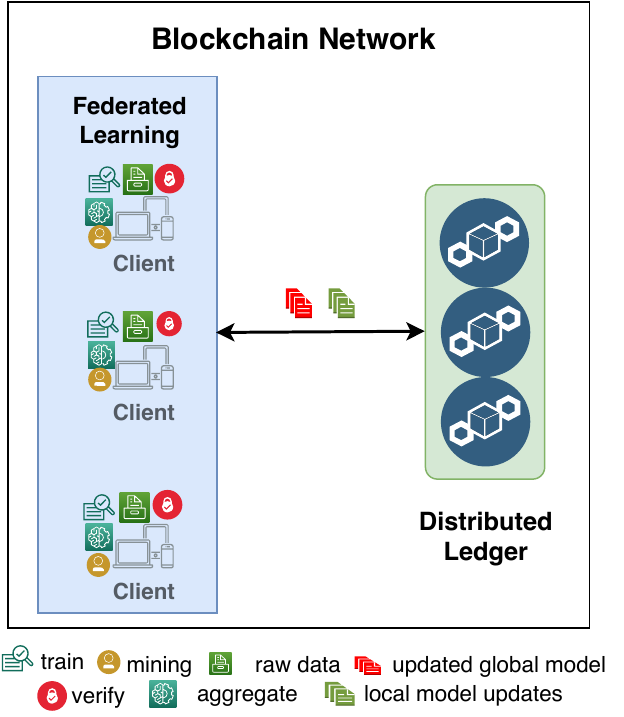}
\caption{The topology of fully coupled BCFL.}
\label{fig:fully}
\end{figure}

\begin{itemize}
    \item Clients collect data and train the models locally.
    \item Local model updates are verified by the (selected) clients.
    \item Verified local updates are collected by (selected) clients and then the global model will be updated.
    \item New block which stores the verified model updates is added into the distributed ledger. 
    \item According to incentive mechanism, rewards will be distributed to participates.
\end{itemize}

FuC-BCFL has been mentioned in various studies. In \cite{Kim2019}, clients of FL are edge sides which can sensor data and provide computing powers, and they are responsible for data collecting and data training.
The blockchain in that framework works as the distributed ledger to record the training data as well. In that system, the integrity of the raw data is protected and the malicious clients are prevented.

The work in \cite{Hua2020} proposes a FL system based on blockchain, where all participates competed to generate new blocks, and then the winner will collect the model parameters and update them into the blockchain. Since no raw data is shared during the training process, the system can preserve the data privacy in a secure manner. 

The FL platform with blockchain is designed in \cite{Toyoda2019}, assuming all the participates can work rationally under the competition incentive mechanism.
This platform can deal with any kind of raw data such as texts, audio, images, etc. Before local model updates are uploaded, several workers will be selected to go through the security procedure under the smart contract to choose the valid data. 

BAFELE is a blockchained FL framework which is central aggregator free and thus decentralized \cite{Ramanan2019}. By delineating the FL mechanism into various rounds and collecting local model updates and then updating the global model, BAFELE can achieve the same model training performance as the conventional FL model. Meanwhile, it costs less computational resources.

From the above discussions, we can conclude the following merits and demerits of FuC-BCFL framework.

\textbf{Merits of FuC-BCFL:} 
\begin{itemize}
    \item The single-point-failure can be avoided effectively as the framework is decentralized and every node has an copy of the distributed ledger.
    \item No data are required to transfer to any central server, avoiding the data privacy leakage and reducing communication cost.
\end{itemize}

\textbf{Demerits of FuC-BCFL:}
\begin{itemize}
    \item More computational resources are required because the operations of both blockchain and FL are running on the same network. Clients not only perform local training, but also integrate the global model.
    \item The communication bandwidth of blockchain network is limited, so the latency of communication could be a challenge to FuC-BCFL.
\end{itemize}

\subsubsection{Flexibly Coupled BCFL}

\begin{figure}[h]
\centering
\includegraphics[width=8cm]{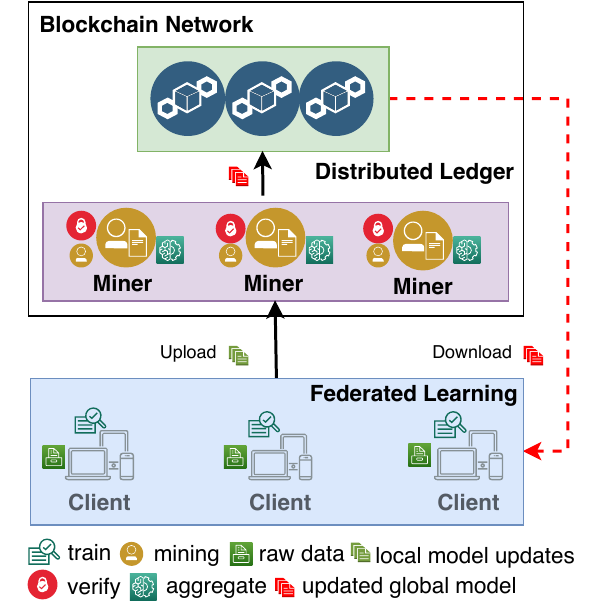}
\caption{The topology of flexibly coupled BCFL.}
\label{fig:flexibly}
\end{figure}

We proposed the flexibly coupled blockchained FL model (FlC-BCFL) when blockchain and FL system are in distinct networks. It means that the clients of FL are not the nodes of blockchain (miners). The topology of the flexibly coupled BCFL is shown in Fig. \ref{fig:flexibly}. From the topology we can see that clients are responsible for local data collecting and training, while local model updates verification will be done by miners on blockchain. FL can also manipulate blockchain to store the model updates, and the miners on blockchain can also aggregate the global model, allowing the central aggregator free in that system. 

\begin{itemize}
    \item Clients collect local data and train the local models, and then upload local model updates to blockchain.
    \item Miners on blockchain perform verification mechanism and only the validated updates can be used to update the global model.
    \item After the global model is aggregated, all the data will be stored on distributed ledger.
    \item Rewards are allocated to participates according to their performances.
\end{itemize}


In \cite{Ma2020}, a reliable and self-motivated FlC-BCFL system is illustrated, which designs a smart contract to publish task and calculate the global model. Nodes train models locally, while miners aggregate the global model on the blockchain. Miners update the global model according to the algorithm defined in smart contract. 

In \cite{Lu2020}, blockchain is used to aggregate the global model and the FL process is executing locally to get local model updates. All the base stations as miners on blockchain execute the global model aggregation process,  while the work in \cite{Lu2020b} leverages one Macro Base Station to integrate the global model. 

Kim et al \cite{kim2019blockchain} proposed BlockFL to exchange and verify local model updates on blockchain. BlockFL focus on removing the central aggregator of FL model, making it decentralized. Miners associated with clients are randomly selected, and local model updates are cross verified among miners. In that paper, the latency of communication on BlockFL is analyzed. 

Reference \cite{Pokhrel2020a} adopts a similar framework with BlockFL which combines autonomous vehicles and miners. The uniform random vehicle-miner association scheme is proposed in that framework, ensuring all the participates can be trusted. To prevent privacy leakage on the Internet of Things devices, the model in \cite{Zhao2020} is composed by manufactures, customers and blockchain. Manufactures establish the learning task and gain the final global model, and customers provide their computational powers to train local models, and at the same time, the blockchain verify and store the model updates. 

We summarise the merits and demerits of flexibly coupled BCFL as below.

\textbf{Merits of FlC-BCFL:}
\begin{itemize}
    \item FL and blockchain are running on different networks and devices, reducing communication pressure and latency.
    \item The raw data remains on the clients, reducing the risk of data leakage caused by malicious attacks on the blockchain network.
    \item Blockchain can provide data sharing for FL, which is more efficient than conventionally FL.
\end{itemize}

\textbf{Demerits of FlC-BCFL:}
\begin{itemize}
    \item Blockchain and FL belong to two different systems, so it is hard to coordinate the management of them.
    \item Single-point-failure still occurs when the central aggregator remains.
\end{itemize}

\subsubsection{Loosely Coupled BCFL}
In \cite{Kang2020a} and \cite{UrRehman2020}, reputation as a crucial criteria is introduced to measure the reliability and trustworthiness of the participates in blockchained FL system. Blockchain in loosely coupled BCFL framework (LoC-BCFL) is used to verify model updates and manage the reputation of participates, and only the reputation related data can remain on distributed ledger. Verification of the updates and reputation management are a part of incentive mechanisms to ensure the participates can behave honestly. We describe the framework of loosely coupled BCFL as shown in Figure 3-3. The workflow of loosely coupled BCFL is as follows:

\begin{figure}[h]
\centering
\includegraphics[width=8cm]{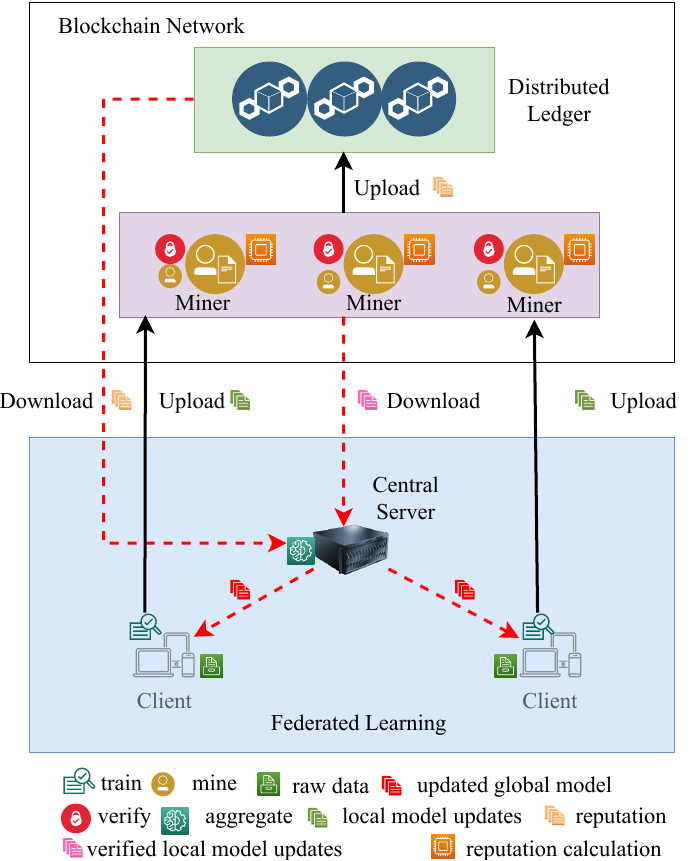}
\caption{The topology of loosely coupled BCFL.}
\label{fig:loosely}
\end{figure}

\begin{itemize}
    \item Clients train models locally and upload  local model updates to blockchain.
    \item Miners verify local model updates, then generate reputation opinions for the clients.
    \item Miners compete to generate new block which contains the reputation related data, and the new block will be added into the distributed ledger.
    \item Aggregator collects the verified updates and then execute the global model aggregation algorithm.
    \item Rewards and penalties are depended on the reputation opinions of clients.
    
\end{itemize}


In \cite{Kang2020a}, reputation and contract theory are combined to support the incentive mechanism for FL. The reputation is calculated by the task publisher according to historical reputation records in that system. The reputation opinions are stored on reputation blockchain after the selected workers finish their proof work. Then the selected workers can start FL process. After training the model locally, workers upload local model updates to task publisher for verification and global aggregation. In that system, reputation is manipulated to choose qualified devices as the workers to conduct federating learning.

In \cite{UrRehman2020}, a reputation-aware fine-gained FL system is proposed to establish a trustworthy computational environment for mobile edges. Reputation of each participate is calculated by public blockchain and smart contracts. The details of other related literature can be fond on Table 3-1.

\textbf{Merits of LoC-BCFL:}
\begin{itemize}
    \item Blockchain and FL are completely independent, and FL retains its data better on its members than the previous two architectures.
    \item The reputation management mechanism enables better management of participants, ensures the quality of data submitted during model training, and improves the accuracy of the model; it also prevents malicious participants from attacking the system.
\end{itemize}

\textbf{Demerits of LoC-BCFL:}
\begin{itemize}
    \item Blockchain is rarely involved in the FL process and only responsible for verification and reputation management, thus the FL model is not decentralized, the risks suck as privacy data leakage, and single-point-failure still exist.
    \item Maintaining blockchain and FL independently will result in inefficient utilization of resources.
\end{itemize}

\subsubsection{Summerized Lessons}
We classify BCFL frameworks into three categories as mentioned above.What's more, we exploit their topology, workflow, merits and demerits. To better understand the characteristics of different structures of BCFL, we summarize the lessons we learned from the above discussions.

\begin{itemize}
    \item We can design different BCFL structures according to specific demands. If the system needs to be aggregator free, then the FuC-BCFL framework is recommended. The FlC-BCFL is suggested when the FL network is not appropriate to run on blockchain network and needs blockchain to assist its learning process for higher model accuracy or data sharing. We can also manipulate blockchain to do reputation management to restrain the behaviors of participates, in this situation, the LoC-BCFL will be a good choice. 
    
    \item Despite the fact that we can classify the structures and they can exhibit different properties, it is currently rather challenging to indicate which structure of the BCFL is the safest and most reliable. We argue that the safety and reliability of BCFL should be evaluated in the perspective of specific computing needs and environments.
    \item Resource constraint and communication latency are impediments to the efficient operation of BCFL and must be addressed regardless of the architecture. 
\end{itemize}

\subsection{Blockchain Types in BCFL}

In this subsection, we will analyze two kinds of blockchain used to assist FL system: public blockchain and permissioned blockchain. We are going to introduce their properties, related works, merits and demerits, respectively. To summarize this subsection, some learned lessons are provided as well.

\subsubsection{Public Chain}
Public blockchain is widely used in blockchained FL system since it's decentralized and transparent. Nodes on public blockchain can be any devices which are willing and have enough powers to take part in the learning process without further certification. 

Reference \cite{Ma2020} proposes a FL system named BC-FL which runs on a public blockchain network. Training nodes and miners can engage in the system without permission and work together to train a global model. Miners on that network take Proof of Work (PoW) as their verification consensus to generate new block. The BlockFL model advocated in \cite{Kim2020} manipulates public blockchain to verify model updates, and the miners are any devices which can provide with sufficient computational powers. Miners compete to complete the PoW, and then the newly generated block will be added to distributed ledger.  To attract more vehicles and base stations to provide data and computational resources, the FL system runs on a public blockchain network \cite{Chai2020}. Proof of knowledge (PoK), a lightweight consensus which combines machine learning with blockchain consensus to avoid complicated computation, is illustrated in that system. In the above models, lowering the barrier to engagement enables more computational resources and more data, however, there are still invalid data and malicious nodes due to less discussion towards misbehaviour detection.

To tackle the security issues of public blockchain used for FL, researchers manipulated protocol designs to prevent misbehaviour from malicious workers to ensure the quality of the learned model. BlockFlow in \cite{Mugunthan2020} is a FL system aided by a public blockchain, however, in order to avoid the malicious clients, the system requires every participate to evaluate each other. Subsequently, a scoring procedure which maintained in smart contract conduct is implemented to reflect the training performance of the clients. By doing these, clients are encouraged to provide high-quality training behaviours to that system. The model in \cite{Toyoda2019} indicates a generic full-fledged protocol to improve the reliability of FL system via permissionless blockchain. Workers are not likely to sabotage the learning process because the competitive model update methodology is designed. 

Following are the merits and demerits of public blockchain in BCFL.

\textbf{Merits of public blockchain in BCFL:}
\begin{itemize}
    \item More data resources and computational powers can be attracted to collaboratively train a common model , thus large scale FL task can be realized.
    \item Public blockchain is totally decentralized and transparent, thus the learning process is traceable and auditable.
\end{itemize}

\textbf{Demerits of public blockchain in BCFL:}

\begin{itemize}
    \item Opening to all devices can lead to difficulties to hinder the law-quality data and malicious behaviours.
    \item Public blockchain in BCFL generally requires complex consensus, e.g. PoW and PoS, to validate model updates and to create new blocks, causing a significant consumption of computing resources.
\end{itemize}

\subsubsection{Permissioned Chain}

In contrast to public blockchain, permissioned blockchain is only available to authorised clients. In the BCFL system, before the devices are registered in the FL, they will be selected based on their computational resources, participation willingness, and historical performances.  

The current research about permissioned blockchain used in BCFL is mainly focus on node selection, i.e., which node can be a part of that chain to continue the learning process. The devices that are intended to be included are usually evaluated before the blockchain starts to operate. In addition, the devices stay or leave at the end of the training depending on their performance. In \cite{Li2020c}, alliance chain is leveraged to enable authority, i.e., nodes management, gradients validation and block generation. Committee Consensus Mechanism (CCM) is designed to validate gradients. The committee is composed by a few honest nodes, and they are responsible for charging the verification process. CCM requires less computational resources than PoW and can perform as a secure and reliable consensus mechanism. Reference \cite{Bao2019} introduces FLChain to settle a reliable and auditable FL ecosystem.  Trainers registered on the blockchain are the entities who are willing to get involved into the training process. Before local model training, miners will be selected according to their reliability and motivation. Malicious trainer's misbehavior will be detected and punished by the authority in FLChain. In \cite{Lu2020c}, FL and permissioned blockchain are integrated. End IoT devices, i.e., base stations and road side unites, are called super nodes on that chain. The local model is trained by committee parties, which are those related registered devices can meet the request of data sharing. Meanwhile, permissioned blockchain remains the data sharing records for audit. 


The following is about the merits and demerits of the permissioned blockchain as applied to BCFL systems.

\textbf{Merits of permissioned blockchain in BCFL:}
\begin{itemize}
    \item The permissioned blockchain offers a platform for a light-weight consensus protocol, which reduces resource consumption while keeping the system secure.
    \item The exposure of the system to malicious attacks is reduced by excluding unauthorised devices from the training of the model.
    \item The performances of authorised nodes can be constrained to guarantee the accuracy of the model, due to the evaluation scheme that usually exists within the permissioned blockchain.
\end{itemize}

\textbf{Demerits of permissioned blockchain in BCFL:}
\begin{itemize}
    \item Less attractive to devices and computing resources than public blockchain.
    \item Reduced applicability of the system due to threshold of access by users.
\end{itemize}

\subsubsection{Summerized Lessons}

When designing a BCFL system, it is necessary to decide which type of blockchain to use. By using different types of blockchain, we can design different BCFL frameworks which satisfy various situations.
With the above discussion we will conclude with some lessons about blockchain types.

\begin{itemize}
    \item We argue that the types of blockchain intrinsically determine the number as well as the quality of the BCFL system's users. Due to the fact that more computing resources and more participants are required in some computing environments, the task of establishing FL on the public chain can be chosen. However, if the training of a model for FL needs to be implemented on a small scale, the permissioned blockchain can be chosen.
    \item The public blockchain and the commissioned blockchain can be used in conjunction with each other, and in \cite{Ramanan2019,desai2021blockfla,fan2020hybrid}, they assist in the training of the model by providing a distinct role for the BCFL system respectively.
\end{itemize}

\subsection{Learning Devices in BCFL}
In this part we will explore the devices in BCFL system, i.e. on which devices FL will run. We argue that, based on the current literature, BCFL can be used for either end devices, such as mobile phones and smart cars, which can sense external data, or edge nodes \cite{Wang2020}, such as base stations, routers and other devices with high computing powers. In the following content, we will discuss the deployment of BCFL on end devices and edge nodes respectively.


\subsubsection{End Devices}

Mobile devices such as mobile phones and automated vehicles generally have computing capabilities, in order to improve computing efficiency, on-device machine learning is used. On-device machine learning requires more data than single device's local data, and the data sharing in devices is necessary \cite{konevcny2016federated,mcmahan2017communication}. FL as a technique for distributed learning, is designed to address the above mentioned issues. The end devices gather external data and train it locally when they are involved in FL. Raw data is not transferred to the server, but only local model updates to the aggregator. When blockchain is used in the above situation, it usually serves to provide decentralized property or as distributed ledgers for FL. This not only guarantees the data privacy of the end devices, but also improves the security of the entire system.

For now, end devices are looking for convenience and intelligence, so it is inevitable that some storage capacity and computing capability is constrained. Current research focuses on the issues when BCFL is used for on-devices, such as communication delays, security leakage, and computing resource allocation. In \cite{Kim2019}, on-device blockchained FL model is proposed. This paper focuses on data exchanges and verification, and arguers that end-to-end latency is an obstacle of BCFL and adjusting the blockchain generation rate could be helpful. However, the limitation of computing capabilities regarding end device is not mentioned. The model in \cite{Hieu2020} considers the above issues and designs a deep reinforcement learning methodology to help machine learning model owner to make the optimal decisions to reduce transmission delay and arrangement the energy consumption. Reference \cite{Lu2020c} leverages blockchain to prevent privacy leakage to secure the data sharing process of the distributed devices. Numerical results shows that the proposed data sharing scheme performs accurately and effectively. 

The merits and demerits of applying BCFL on end devices are listed as below.

\textbf{Merits of on-device based BCFL:}
\begin{itemize}
    \item Raw data is not required to be transferred to any other devices, reducing the resources consumed by data transmission, while data security is ensured.
    \item The usage of end devices is widespread, thus attracting more users and generating more data for model training.
\end{itemize}

\textbf{Demerits of on-device based BCFL:}
\begin{itemize}
    \item End devices have limited computing, storage and communication capacity to undertake complex local computing.
    \item An end device is not only responsible for local data collection, model training and data storage, but also for providing the resources to keep the blockchain network running, which may result in the device being unable to do other tasks properly.
\end{itemize}

\subsubsection{Edge Nodes}
In \cite{wang2019adaptive,wang2019edge,qian2019privacy}, FL technology is used to support edge computing. In conventional edge computing scheme, raw data is sent to nearby edge node, which can be considered as the central server where raw data will be proceed. Despite FL can avoid the transmission of raw data by training the raw data locally and uploading the model updates to the edge node or central server, the risks of FL itself such as single-point-failure and privacy leakage still remain. By leveraging blockchain to support the FL based edge computing, the whole system can be more secure and reliable. In the system which combines edge computing, FL and blockchain, all the end devices collect the raw data and then sent them to the nearby edge nodes for model training; blockchain provides data verification and data sharing for edge nodes; and the verified model updates will be transmitted to central server for global model aggregation. 

Reference \cite{Cui2020} introduces a system named CREAT, which adopts blockchain to help the edge computing to catch content during the FL process. IoT devices transfer collected data to blockchain, and each edge node downloads the data and then computes the gradients independently. The original purpose of applying FL model to edge computing is to ensure edge nodes can collaboratively learn the features of users and files so that the catch hit rate can be promoted by predicting popular files. Blockchain is incorporated to secure the data transmission and sharing. In \cite{fan2020hybrid}, edge nodes equipped with computational powers and storage can receive the data from end devices, and train the deep learning model collaboratively. Public blockchain and permissioned blockchain provide the collaboratively and auction mechanism to the FL system, respectively. 

Here are the merits and demerits of learning devices of BCFL.

\textbf{Merits of edge nodes based BCFL:}
\begin{itemize}
    \item Edge nodes based BCFL are able to provide sufficient storage capacity and computational resources.
    \item Edge computing can be more secure and reliable, and its application is wider. 
\end{itemize}

\textbf{Demerits of edge nodes based BCFL:}
\begin{itemize}
    \item Raw data needs to be transferred, which reduces security and increases the consumption of the resources required for transmission.
    \item The distribution of edge nodes is not as widespread as that of end devices, which may limit the application of BCFL.
\end{itemize}


\subsubsection{Summerized Lessons}

In this section, we explore the scenario when BCFL is deployed on end devices and edge nodes respectively. Besides inconsistent workflows, different devices can affect the overall performance of the system. We will conclude this subsection in the following.

\begin{itemize}
    \item The overall difference between the ways in which end devices and edge nodes are involved in a BCFL system is that the former keeps raw data locally, while the latter needs to collect raw data from multiple devices.
    \item From the blockchain level, some of the models' blockchains are maintained via edge nodes, while others are maintained via end devices. The devices on blockchain network may not participate in FL training process, we have noticed this in Section \ref{sec:framwork}.
\end{itemize}

\section{Functions of BCFL}\label{sec:functions}


In this section, we investigate some specific functions of BCFL with the perspective regarding its workflow, including verification of model updates, aggregation of global model, utilization of the distributed ledger, and incentive mechanism.

\subsection{Verification of Model Updates}\label{sec:verification of Model Updates}
To train a well performed global model, FL needs to ensure that all the devices engaged in the model training process work honestly and provide reliable data. This problem is not well tackled in traditional FL models. To address this issue, we can take advantage of blockchain to verify the submitted data,  excluding the dishonest and unreliable data.

\subsubsection{Verification Protocol}\label{sec:verification protocol}

In each round, the local devices transmit the trained local model updates to the miners for further validation (no data transfer is required in FuC-BCFL framework). Therefore, a suitable validation mechanism needs to be designed to verify the validity of the data and to reduce the time and resources consumed.

Current research places significant emphasis on verification mechanisms. The work in \cite{Kang2020a} proposes a Proof of Verifying (PoV) consensus to ensure the uploaded local model updates are valid before the global model aggregation. The main idea of PoV is to prepare the testing dataset in advance and to set a threshold for accuracy. According to PoV, the reliable testing dataset provided by the task publisher will be prepared on the blockchain, and then the miners utilize this dataset to verify the uploaded updates. The qualified updates are selected based on a given accuracy threshold and put into blocks as transactions. The threshold can be determined empirically, but the selection of testing dataset is a challenge because it is hard to use the previous data for valuation once a new learning environment is situated. 


The verification process in \cite{Li2020c} is similar to the PoV mentioned above, and the miners in a committee are responsible for verifying the updates and scoring them, while the details of how to score the updates are not mentioned. 


Reference \cite{Cui2020} designs smart contracts to verify the transactions storing local model updates. The whole process requires the randomly selected consortium members to vote whether the updates are reliable, and the decision is based on the number of received votes. Although randomly selected members are required to participate in the voting, it is hard to show that this avoids the influence of subjectivity, so more evidence is needed to support this methodology. 

The work in \cite{Lu2020a} designs a two-stage verification scheme, which manipulates cumulatively calculated reputations based on the accuracy of the updates and nodes on the blockchain to evaluate the quality of the transactions. 

Although the importance of validation mechanisms is mentioned in some studies, no specific descriptions of the workflows are provided \cite{kim2019blockchain, Pokhrel2020a}.







\subsubsection{Summerized Lessons}

\begin{itemize}
    \item 
    The verification mechanism can be designed in various forms, but it is more common to filter the updates before conducting model aggregation to avoid unreliable data from affecting the global model. Of course, it is also possible to manage the updates through the feedback after model aggregation.

    \item By validating the updates, the verification mechanism can not only filter out the unreliable data, but can also constraint the dishonest behaviors of data providers. In addition, the results of verification can also be used for the later guidance of rewards allocation.

    \item Based on our research, although researchers realized the importance of  the verification of model updates before aggregating them, studies about the design of effective validation mechanisms are still lacking.

\end{itemize}

\subsection{Aggregation of Global Model}\label{aggregation of global model}
The basic idea of FL is to distribute model training tasks to numerous local devices and then to integrate the local models through a central aggregator. Therefore, model integration is a crucial component of the FL process. In the following section, we will explore how to utilize blockchain technology to assist the aggregation of global model for FL. Based on our investigation of current research, our analysis will focus on the members who are engaged in model integration in the BCFL framework. In Section 3.A, we discussed the architectures of BCFL, and we found that in some BCFL models, the central aggregator are still remained since the blockchain and FL are coupled in different ways \cite{passerat2019blockchain,Lu2020b}. In the following content, we are not going to discuss this kind of situation because we are rather interested in knowing how decentralized model integration is enabled via the application of blockchain. 



\subsubsection{Selected Blockchain Nodes}\label{sec:selected}
In some models, after local model updates are verified by the nodes on the blockchain, only the selected nodes participate in global model integration. Those selected nodes are usually well equipped with enough computational resources or have good historical performance records.

In \cite{Li2020c}, the authors propose a committee consensus mechanism to verify local model updates and then aggregate the global model. They argue that the election of the committee is crucial to performance of the global model, and they also introduce three kinds of committee election methodologies, including random election, sorting by score, and multi-factor optimization. The experimental results show that the model under this mechanism can obtain similar performance as the conventional FL model. 

In \cite{Lu2020c}, the committee nodes are responsible for model training and aggregation, which are selected according to their registration records. This kind of election of committee lacks the evaluation of data provider's reliability, leaving the quality of raw data uncertain.

By selecting some nodes to participate in the model integration, on the one hand, it can avoid the existence of a central node and achieve decentralization; on the other hand, the selected nodes are usually more reliable, and the overall resource consumption can be reduced by implementing them to complete the model aggregation.

\subsubsection{All Blockchain Nodes}


When all the data providers or miners are independently involved in the aggregation of the model, such a framework is decentralized and avoids any authority center completely. This is the most commonly used framework for applying blockchain to FL.

Fully decentralized global model aggregation is usually done by miners or data providers on the blockchain, i.e., local devices. In flexibly coupled BCFL models, miners and data providers are not the same, and each miner aggregates the global model via aggregation algorithms after finishing the verification of local data updates \cite{Ma2020,Kim2019}. While in the fully coupled BCFL framework, the local devices are usually the miners, so they not only collect the data and then train the local models, they also verify the updates and calculate the global model\cite{Toyoda2019, Preuveneers2018}.

By replacing the central aggregator with the blockchain, the task of model integration is delegated to nodes on the blockchain, which can be miners or data providers, depending on the different coupling framework. In that case, the BCFL can be completely decentralized that every node can participate in model aggregation, avoiding single-point-failure effectively.

\subsubsection{Summerized Lessons}
\begin{itemize}
    \item Blockchain allows FL to modify the process of model aggregation, leaving central aggregator unnecessary.
    \item No matter the global model is computed by partial nodes or all nodes, the integration of the model can be effectively decentralized.
\end{itemize}

\subsection{Utilization of the Distributed Ledger}

In the conventional FL model introduced by Google\cite{Konecny2016a}, the raw data are kept on the local devices while local model updates shall upload to the central aggregator. With the help of blockchain technology, FL can work effectively without the central aggregator. When the miners finish the verification work, the new block will be generated and added to a blockchain where the validated local model updates and the aggregated global model are stored\cite{kim2019blockchain}. In this process, blockchain works as the distributed ledger, which stores the model updates and provides an accessible platform for all the qualified participates to retrieval the data. In this subsection, we will discuss two aspects of blockchain as the distributed ledger in the BCFL model: data storage and data sharing. 

\subsubsection{Data Storage}
In conventional FL, local model updates are generally transferred to the central aggregator and then stored, requiring more transferring cost and storage capacity. By incorporating blockchain for assisting FL, the problem of data storage in the training process can be effectively ameliorated. To some extent, blockchain is a distributed ledger that can provide a secure, traceable, and immutable way to store data. All the data related to model training, including local model updates, global model updates and reputation of the participates, is treated as the transactions of blockchain and is needed to be verified by the miners. First, only the validated data can be recorded in the newly generated block, and then the block will be added to a blockchain. By this design, the data in distributed ledger is traceable and immutable, which means once the transaction is added to the blockchain, it is nearly impossible for any device to change the records. 

Current research is less concerned about the concrete structure of the blockchain in BCFL. The work in \cite{Sharma2020} describes details of the structure of the blocks in the blockchain used for FL. A block consists of a block header, which contains information such as model ID, data ID, timestamp and data types, and a block body, which holds model updates. 

In \cite{Li2020c}, the recommended system chooses the alliance blockchain to store the data, allowing only the authorized participates to access to the ledger. The blocks on that blockchain are varied: one called model block is used to store the global model for each round; and the other one is named update block, which is implemented to store local model updates and other learning information such as address of devices and update scores. 




\subsubsection{Data Sharing}
In Google's conventional FL model, only the central aggregator can get the updates from the devices \cite{Konecny2016,Konecny2016a}, while in blockchained FL model, all the qualified participates can access to the blockchain to retrieval and share the data to support model training. Blockchain provides a data sharing platform for FL to train a machine learning model with better generalization capability. What's more, the data shared during the training process is local model updates and other related data (e.g., reputation, IP address, timestamp and so on.) rather than the raw data \cite{UrRehman2020,  Awan2019,Ma2020}. In this case, the data privacy can be well protected and the efficiency of model training can be improved. 

Some research focuses on designing the scheme of data sharing based on blockchained FL \cite{Lu2020a,Lu2020c}. For example, reference \cite{Lu2020c} builds a permissioned blockchained FL environment to share the data among distributed industrial IoT devices. In permissioned blockchain network, there are two kinds of transactions should be proceed: data retrieval and data sharing. The local devices communicate through the blockchain, which can ensure the security of data transmission. The super nodes on the permissioned blockchain, i.e., routers, base stations, and other facilities with strong computing powers, keep the records of the local devices of the IoT after being encrypted. In addition, in order to improve the efficiency of data retrieval and model training, local devices with the same data type are grouped in a community. In each committee, the ID information of each participant is public. By this design, the data can be shared in an efficient and secure way. The authors argue that the encryption methodology for data sharing can't avoid data leakage, thus they design a request and reply protocol between the data requester and the permissioned blockchain. After the requester sends a request for data sharing, the blockchain members will first check whether there are already records that match the request, and return the result directly if there are; if not, they will train the model through the relevant committee nodes and finally return the result. In this model, blockchain provides a platform to store data and retrieve it securely. However, since this data sharing framework involves storing the model for retrieval in advance and keeping the data of local devices through super nodes, further research is needed to investigate whether it can effectively prevent external attacks.

\subsubsection{Summerized Lessons}
\begin{itemize}
    \item From the perspective of learning process, blockchain provides distributed data storage and public data sharing for FL. Instead of storing the data generated during the learning process in the central aggregator, FL only needs to store this data through the blockchain, allowing the relevant data freely available to all authenticated participants.
    
    \item From the perspective of data security, the blockchain itself can be seen as a distributed ledger with characteristics such as immutability, auditability and decentralization. Blockchain can record all necessary data and also prevent malicious nodes from altering it. And only authenticated participants can access the data related to FL, preventing the privacy leakage.
    
\end{itemize}





























\subsection{Incentive Mechanism}
This subsection will discuss how the incentive mechanism in BCFL ensure that participants work honestly according to the protocol, ensuring the final trained global model reliable.

\subsubsection{Incentive Mechanism Design}
FL offers a distributed computing solution for machine learning. However, traditional federation learning models cannot guarantee that all participating clients are reliable. Blockchain can address this issue by distributing the corresponding rewards to nodes that have contributed in the generation of blocks based on their contributions. In this way, by incorporating a blockchain into the FL model and rewarding the participants (local devices and miners) according to a certain scheme, participants can be motivated to provide reliable training data. In addition, the incentive mechanism can also penalize dishonest nodes, filtering out the malicious participates.

Incentive mechanisms have been emphasized in the existing studies. The work in \cite{weng2019deepchain} designs a payment-based incentive mechanism to encourage participates to collaboratively train a deep learning model. Two properties of that incentive mechanism are introduced, i.e., \textit{compatibility} and \textit{liveness}. \textit{Compatibility} assumes that all the participates can get maximum rewards based on their contributions, and liveness means that all the participates have the willingness to update both the local model and global model. After the final global model is updated, the rewards will distribute to local devices and miners according to their contributions.

In \cite{Toyoda2019}, repeated competition is implemented to motivate the workers to obey the rules of the protocol in order to obtain the maximum profits. The basic idea is to introduce a mechanism for workers to compete for the opportunity to update models at each training round and to constrain their subsequent performance through a voting scheme. The distribution of the returns will be determined by sorting the records of the votes.

Reference \cite{desai2021blockfla} argues that monetary is most popular incentive for participates in BCFL and illustrates a penalty scheme which requires each participate to deposit a certain amount of cryptocurrency on the blockchain. When the global model is well trained, the deposits will return to the participates, and additional rewards are distributed to encourage honest behavior. The rewards are determined by the average time participants spent on submitting trained data, with faster submissions being awarded more. On the contrary, if one participate is found being dishonest, then it's deposits will lost.

In other studies, there are distributions of returns based on calculating the contribution of participants in model training \cite{Li2020c, Zhang2020a}, and the management of participants in these studies is based on reputations \cite{kang2019incentive, UrRehman2020, Zhao2020}.




\subsubsection{Summerized Lessons}

\begin{itemize}
    \item Incorporating an incentive mechanism into the FL model to give participants certain rewards can effectively regulate and discipline their behaviors and can encourage participants to provide reliable training data.
    
    \item  Current research lacks in-depth study on how to allocate rewards. On the one hand, a decentralized evaluation system needs to be designed; on the other hand, some defects of the blockchain itself should be taken into account when designing incentive mechanism.
\end{itemize}

\section{Applications of BCFL}\label{sec:applications}
FL and blockchain are already being applied in many fields. Instead of exploring the real-life applications of both separately, this subsection will investigate the applications of BCFL. According to the current research, BCFL has been initially applied in the fields of Internet of Things, smart city, financial payment, and healthcare, etc. Even though these research are all based on specific usage environments to apply BCFL, there is no general framework yet.

\subsection{BCFL for IoT}

In IoT area, devices are decentralized, so model training on them requires timely and secure data and strong model generalization capability. FL in the Internet of Things(IoT) can collaboratively train a global model by numerous devices, avoiding the leakage of data privacy \cite{Du2020,Khan2020}. However, FL itself has several deficiencies (e.g., single-point-failure and lack of incentives), and blockchain can improve the security of model training for IoT devices.

Research on the applications of BCFL in the IoT domain focuses on data security, resource planning, communication, and failure detection, with the aim of enabling IoT devices to jointly train a model with good performance. 

The work in \cite{Lu2020a} introduces a BCFL model to protect the privacy in Internet of Vehicles (IoV), and in \cite{Lu2020}, communication efficiency and resource limitation in IoT devices based on the BCFL framework are investigated. In industrial IoT (IIoT), the data heterogeneity in failure detection challenges the reliability of the whole system. The work in \cite{zhang2021bc} utilizes BCFL to design a secure data transmission method in IIoT environment. The BCFL frameworks are researched in unmanned aerial vehicles (UAV) to protect privacy and improve efficiency\cite{feng2021blockchain, aloqaily2021energy}. In \cite{Zhang2020a}, a blockchained FL model is proposed for failure detection in IIoT. First, a FL model is deployed among IoT devices and a central server is set up for model integration; then, data from local devices is stored via blockchain, which also provides incentives. In the aspect of failure detection, a new aggregation algorithm is designed to reduce the impact of data heterogeneity by considering the distance between positive class and negative class in each dataset. 




\subsection{BCFL for Healthcare}

In healthcare area, data of patients is sensitive, so both patients and hospitals are reluctant to share their heal data. FL can help train the model distributively, while the data leakage is the biggest challenge \cite{choudhury2019differential,chen2020fedhealth}. Blockchain can be implemented among patients or hospitals, allowing participates to share data without privacy disclosure. 

Passerat-Palmbach et al. \cite{passerat2019blockchain} point out that the protection of patient privacy constrains researchers from analyzing health data, and existing tools are insufficient to address the issue, so they suggest to use both blockchain and FL for healthcare consortia. In their model, data access, model integration, weight encryption, and auditing of the learning process are emphasized. However, this study is specific to consortia and is not appropriate for most health problems and lacks concrete solutions which can be operated. Polap et al. \cite{polap2021agent}designs a multi-agent architecture based on BCFL for the internet of medical things (IoMT) to protect data privacy.

In contrast to \cite{passerat2019blockchain,polap2021agent}, Kumar et al. \cite{kumar2021blockchain} offer an specific solution for COVID-19 detection via BCFL models. Hospitals train the local model based on their own private data and share only the weights and gradients, and blockchain records the learning process and related data. Researchers highlight the privacy of patients, and BCFL framework can protect the privacy when the global model is training. That paper builds a secure and decentralized data sharing platform among hospitals, enabling the automatic detection of COVID-19 in a secure manner.

The research used BCFL in healthcare is rare now, but the research direction is promising since a large amount of medical data have to be proceed and BCFL can offer secure learning solution.





\subsection{BCFL for Business and Finance}

Blockchain emerged as the basis for Bitcoin at the beginning, and the explosion of various blockchain-based virtual currencies in recent years in particular has shown the importance
of blockchain as an underlying technology in finance and business. Meanwhile, FL can offer a distributed machine learning framework. Therefore, BCFL can provide secure and decentralized applications for the financial and business fields. 

The most direct application of BCFL in the finance and business field is to provide a monetary payment method. FedCoin, introduced in \cite{Liu2020}, provides a peer-to-peer payment system based blockchain for FL. FedCoin is different to Bitcoin, which depends on PoW, it utilizes the proof of Sharpley (PoSap) to generate new blocks. Such a payment system can be applied to a commercial system based on FL.

In addition, BCFL can be applied in the areas of financial investment, for example, the processing of financial big data. Since data from customers of financial companies is sensitive, customers do not want to disclose their data to the concern of privacy and companies are obliged to keep their customers' confidential. Therefore, companies can deploy BCFL to obtain data and train models to develop more accurate market-oriented financial products, while protecting customer privacy.

\subsection{BCFL for Smart City}

The construction of a smart city requires a large amount of data, and by training these data and getting reasonable models, it can provide better services to citizens. Similar to many machine learning situations, privacy and security have been constraints to the development of smart cities.

BCFL can provide a secure big data training architecture, while offering rewards based on user contributions to motivate users to provide more data. Imagine this scenario, when some institutions need to optimize urban traffic and require multiple devices and users to provide data and collaboratively train models, what kind of model training framework should they use? Traditional machine learning frameworks, including conventional FL framework,  cannot guarantee privacy and provide incentives at the same time. Fortunately, the requirements can be met with BCFL. In \cite{qi2021privacy}, a BCFL framework with differential privacy approach is proposed to predict the traffic flow.

The major advantage of BCFL for smart cities is not only that it can protect privacy and deliver incentives, but also that it can allow more devices to join, adapting to the large number of devices and users in a smart city.





\section{Challenges and Future Research directions}\label{sec:challenges}

While BCFL has many advantages, some challenges that may hinder the operations of the BCFL model cannot be ignored. 
In both FL and blockchain, each of them faces many design challenges. To be specific, FL needs to address issues such as Byzantine attacks, free riders, non-IID data and device selection during model training, and blockchain needs to address problems such as being attacked and transaction pricing. In the BCFL system, some of the problems can be solved by existing methods, for example, Byzantine fault-tolerant FL aggregation algorithms can be designed to defend against poisoning attacks and backdoor attacks \cite{cao2020fltrust,guo2021siren,li2021byzantine,xu2021signguard}.
In this section, we will discuss the deficiencies of BCFL, i.e., the new challenges that the inclusion of blockchain brings to FL.
We argue that a good BCFL model should have high security, high training efficiency and low computational cost. The design of BCFL is a trade-off between these three aspects, and our following analysis will be carried out from them.

\subsection{Privacy and Security}

Security and privacy are of importance to the BCFL model, and although both blockchain and FL have privacy-preserving properties, there are still some issues that may lead to privacy leakage.

\subsubsection{Anonymity}

In the conventional FL model, only the central center knows the sources of local model updates. However, the addresses of clients are public in BCFL, and other clients can obtain the training behaviors based on the public information from blockchain. What's more, clients generally do not communicate with each other addresses are private information. While in BCFL, since there exist identity information such as public addresses, clients may be able to communicate with each other, increasing the risk of collusion among clients.

\subsubsection{Shared Data}

Blockchain stores the blocks which contain the model updates through a chained structure, and all members within the blockchain can access the data from the public distributed ledger as well as download the data. In BCFL, clients can get information about other members from blockchain. In BCFL using public blockchain, since there is no access restriction, information of members may be available to external devices, threatening the security of the whole system. Data sharing can improve the speed of model training and facilitate clients to perform model updates, but the risks associated with data security cannot be ignored.

\subsubsection{Malicious Attack}
In the decentralized BCFL model, there is no authority center to regulate the behaviors of participates, therefore the risks of being attacked by potential malicious participates exist. On the one hand, the attacks may come from blockchain system, such as forking, double spending, and selfish mining. Forking is one of most common attacks launched by attackers, which tries to obtain more profits by replacing the most trusted chain (i.e., longest chain) with an alternative chain. Double spending occurs when a currency is spent twice. Selfish mining attack, also named block withholding attack, happens when an entity validates one block but does not broadcast it to the network. On the other hand, attacks from FL system will hinder the deployment of BCFL, including data poisoning, inference attacks, etc. Malicious users can launch data poisoning attack by utilizing dirty data to train the local models, and then upload the biased local models to the aggregator, leading the parameters of global model inaccurate. Even though the uploaded parameters are encrypted, malicious users can still deduce the real information by analyzing them, so inference attacks may cause the leakage of privacy in FL system. 

Since malicious attacks deteriorate the reliability of BCFL, future research can focus on the combination of the two technologies to reduce the risk of being attacked. For example, reasonable mechanisms can be designed to use blockchain for the selection of users and data.

\subsection{Training Efficiency}

The goal of FL is to train a global model through the collaborative work of multiple devices, not only the accuracy of the global model, but also the time and computational cost consumed by training, should be taken into account.

\subsubsection{Reliability of Data}
Since we cannot guarantee that all participants are honest, it is unreasonable to assume that all data are reliable. We cannot ignore the impact of unreliable data on the global model. At least three measures are considered to improve the reliability of data: 
\begin{itemize}
    \item Perform clients selection before training to exclude potentially dishonest nodes. The impact of the types of blockchain needs to be considered in the selection of clients. When BCFL uses public blockchain, any device can join the training without permission. In this case, the selection scheme can be designed to decide whether allow those devices to continue to participate in the training based on the performance of the clients in the previous round. It requires to evaluate the device performance in single or multiple rounds. In fact, many verification mechanisms have adopted this approach. As for the permissionsed blockchain, since potential devices need to obtain permissions to join the training, malicious attacks and invalid data can be reduced to some extent.

    \item Design efficient verification mechanisms to speed up the processing so as to reduce the time consumption and to improve the accuracy of verification, ensuring only qualified data can participate in model integration. The current research lacks detailed study of the process of the verification mechanism. Devices involved in the verification mechanism need to be considered. No matter data is verified among clients or through miners on the blockchain, privacy needs to be prevented from being leaked. Different verification mechanisms can affect the security of the model.

    \item A reasonable incentive mechanism should be designed to encourage participants to provide truthful data, and penalizing those who are dishonest. Most of the current research focuses on how to distribute rewards, and we believe that innovation can be made from the perspective of punishment. In the blockchain ecosystem, behavior of clients can be constrained by depositing a portion of virtual currency (e.g., bitcoin and ethereum) before training.

\end{itemize}


\subsubsection{Communication Latency}
Communication latency occurs in both FL and blockchain networks, which are also a constraint to the development of these two technologies. Latency analysis has been given enough attention in BCFL, and a number of studies have already proposed solutions, for example, Kim et al. \cite{kim2019blockchain}]suggest reducing the computational difficulty of PoW to lower latency.


\subsubsection{Asynchrony}
During the training process, the time of participates entering and exiting influences the effectiveness of the training. The time to join training can be specified by designing a participant selection mechanism, however, several factors can cause participants to drop out of training early, such as network issues, damaged devices, limited storage space, etc. The above-mentioned problem affects the distribution of rewards apart from the correctness of the global model.

\subsection{Resource Consumption}

Storage consumption, computing consumption and communication consumption compose the main consumption of BCFL.

\subsubsection{Storage Consumption}

In the conventional federation learning model, local model updates are stored on the aggregator, while in BCFL, data is stored through the blockchain. Meanwhile, every clients can also keep a copy of the blockchain locally and update it continuously, increasing the storage cost. For devices with insufficient storage capacity, they may not be able to continue to participate in training as the data stored in the blockchain grows. In addition to the storage of data, it will be a research direction how the clients and miners can efficiently retrieve data on blockchain.

\subsubsection{Computing Consumption}
FL usually requires multiple rounds of iterations to get the final global model, so the cost of model training is usually related to the number of training iterations. The trade-off between model accuracy and cost has been a topic for researchers. Compared with traditional FL models, BCFL requires not only local model training, model aggregation and updating, but also data validation and block generation. These activities consume a large amount of computational resources and increase the cost of training models. In \cite{li2021blockchain}, the authors propose a joint resource allocation mechanism for the clients in FuC-BCFL system, which assume that all the devices have the same computing powers and train the local models with the same iteration steps, making this methodology not applicable in mobile scenario.

In FlC-BCFL model, miners and clients are different devices, and the cost calculation needs to be based on different roles. For miners, a significant amount of computations will be used to run consensus protocols, i.e., mining, which is an arithmetic-intensive process, so light-weight consensus protocols can be designed to reduce the computational difficulty. In addition, the overall cost can be lower by reducing the cost of data verification. For clients, reducing the number of training sessions while ensuring the training quality can reduce the cost.

The training cost of the model is not only related to the accuracy of the global model, but also affects the security of the whole BCFL model. For example, if the difficulty of generating blocks is reduced or the process of verification is simplified, although the computational cost can be reduced, it may lead to security problems. In addition, from the overall perspective of BCFL, the blockchain and FL need to operate in a coordinated manner, and the methods of allocating resources will also affect the computational cost. Therefore, computational cost is a topic that needs to be addressed gradually in future research.

\subsubsection{Communication Consumption}

In BCFL, in addition to the FL process, which requires device-to-device communication, communication in the blockchain system also needs to be taken into account. Since blockchain is a P2P network, once the number of blockchain nodes increases, then the communication cost will increase exponentially. Besides, the size of the amount of data to be shared also affects the communication cost. We argue that the number of communications, the number of devices, and the size of the transmitted data should be considered in future research.

\subsection{Future Research Directions}
Besides some research directions towards the challenges of BCFL mentioned above, we also provide some promising directions in this part.

\subsubsection{BCFL in Mobile Scenarios}
Deploying BCFL in mobile scenario will encounter the following two main challenges: 
\begin{itemize}
    \item How to address the lack of resources, i.e., storage, computation, and communication resources? As mobile devices generally do not have sufficient resources, so offloading computation tasks to the edge server could be an option, but there is an ensuing challenge of privacy leakage. In addition, the selection of BCFL architectures is important. In general, FuC-BCFL is not suitable to be deployed in mobile scenarios since it requires a large amount of resources. 
    \item How to select devices to form the blockchain network and join the FL process? Because mobile devices may be in dynamic status, the network they form is unstable, and the impact of devices joining and exiting needs to be considered, causing increased difficulty in maintaining the BCFL framework.
\end{itemize}
\subsubsection{Byzantine Robust BCFL}
FL and blockchain systems could be exposed to Byzantine attacks respectively. The combination of the two techniques could lead to new attacks that would make existing methods ineffective. We argue that possible attacks against BCFL may come from the data sharing process in BCFL, because BCFL shares local model updates, which may lead to data plagiarism, poisoning attacks, and backdoor attacks. In addition, possible attacks may come from the consensus process on the blockchain, which means that some malicious nodes may control the working process of BCFL, thus making the whole system unstable.
\subsubsection{Fairness of BCFL}

We argue that there are at least two aspects of fairness that can be discussed: i) devices should be fairly selected as FL clients or as miners on blockchain to participate in BCFL; ii) all devices involved in the work of BCFL should fairly receive the corresponding rewards. Note that the so-called fairness does not imply absolute equality, but is based on certain evaluation metrics, such as assigning rewards according to the contributions of the devices.

\subsubsection{Learning-aware BCFL}
We argue it is possible to deploy another mechanism used for learning on BCFL to assist it in attack detection and defense, device selection, assigning rewards, etc. The deployment of such a mechanism, however, requires consideration of resource consumption and privacy protection.

\subsubsection{FL-assisted Blockchain}
Current research focuses on how to use blockchain to assist FL to better accomplish model training, however, whether FL can be used to assist blockchain to reach consensus and maintain sustainability has not been explored. In terms of research difficulty, FL- assisted blockchain is more difficult to be implemented, but we should not ignore the possibility of the existence of such an architecture.

\section{Conclusion}\label{sec:conclusion}
In this paper, a detailed investigation of blockchained FL (BCFL) is provided. We first introduce blockchain and FL respectively. Then we investigate the foundations of BCFL, including the architecture of BCFL, the blockchain on BCFL, and the devices on BCFL. We also analyze four functions of BCFL, i.e.,  verification of model updates, aggregation of global model, utilization of the distributed ledger, and incentive mechanism. After that, we survey the applications of BCFL in real life. Finally, we discuss the existing challenges of BCFL and give the corresponding future research directions.

Blockchain and FL are both emerging technologies, and their combination can efficiently address the security and privacy issues of distributed machine learning. This paper is the first detailed survey on BCFL, and we believe that BCFL will be used more often in the future. We hope our work will bring new ideas for future BCFL research.

\bibliographystyle{unsrt}
\bibliography{reference.bib}

\if()
\begin{IEEEbiography}[{\includegraphics[width=1in,height=1.25in,clip,keepaspectratio]{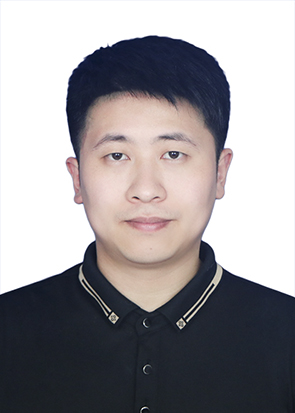}}]{Zhilin Wang} received his B.S. from Nanchang University in 2020. He is currently pursuing his Ph.D. degree of Computer and Information Science In Indiana University-Purdue University Indianapolis (IUPUI). He is a Research Assistant with IUPUI, and he is also a reviewer of 2022 IEEE ICC, IEEE Access, IEEE TPDS, IEEE IoTJ, and JNCA, and I also serve as the TPC member for IEEE ICC 2022 Workshop. His research interests include blockchain, federated learning, edge computing, and optimization theory.
\end{IEEEbiography}

\begin{IEEEbiography}[{\includegraphics[width=1in,height=1.25in,clip,keepaspectratio]{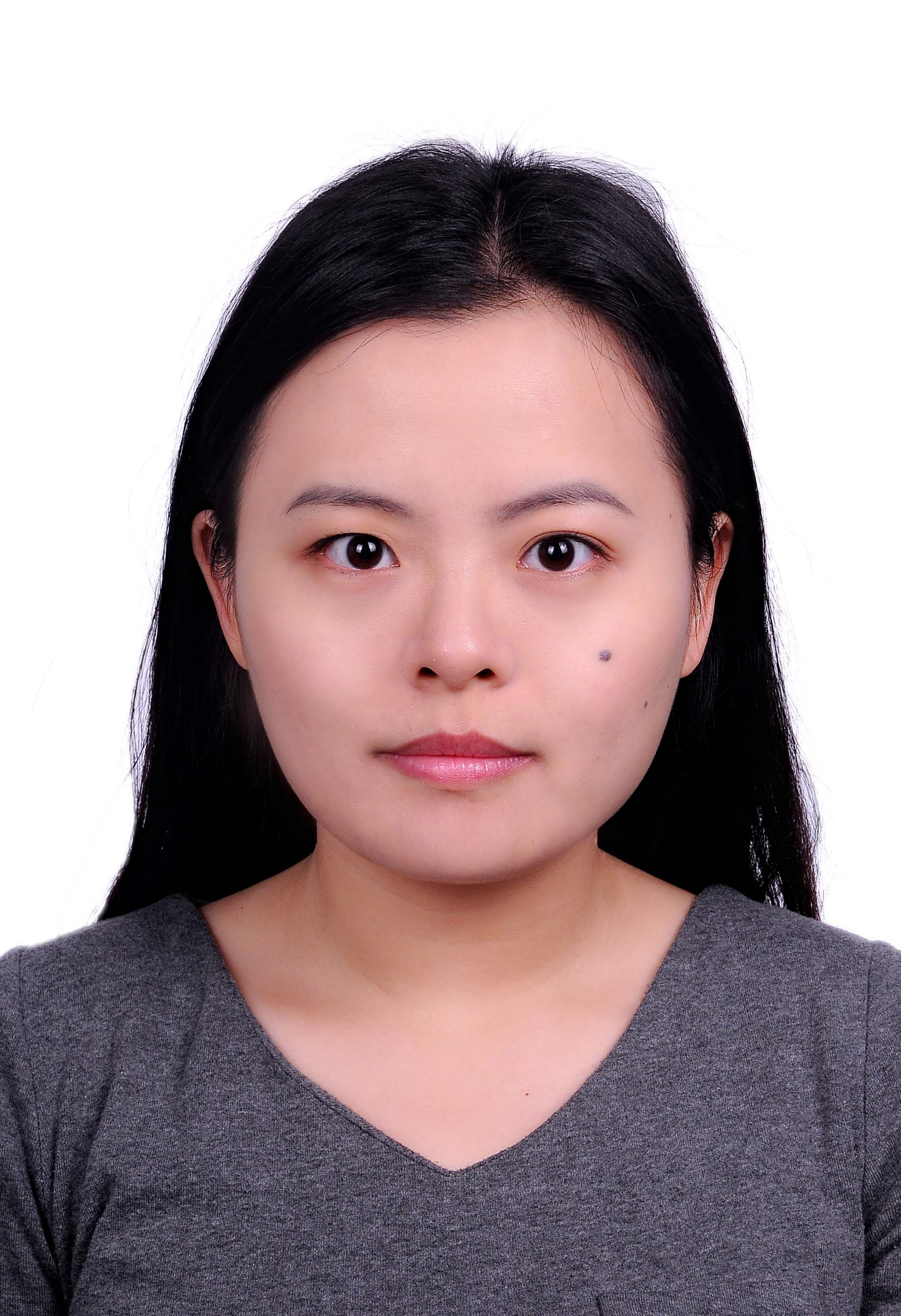}}]{Qin Hu} received her Ph.D. degree in Computer Science from the George Washington University in 2019. She is currently an Assistant Professor with the Department of Computer and Information Science, Indiana University-Purdue University Indianapolis (IUPUI). She has served on the Editorial Board of two journals, the Guest Editor for two journals, the TPC/Publicity Co-chair for several workshops, and the TPC Member for several international conferences. Her research interests include wireless and mobile security, edge computing, blockchain, and crowdsensing.
\end{IEEEbiography}
\fi








\end{document}